\documentclass[journal,onecolumn,10pt,twoside]{IEEEtran}

\ifCLASSINFOpdf
\usepackage[pdftex]{graphicx}
\else
\fi
\usepackage{cite}
\usepackage{epstopdf}
\usepackage{amsmath,amsthm, amssymb}
\usepackage{graphicx}
\usepackage{textcomp}
\usepackage{threeparttable}
\usepackage{bm}
\usepackage{multirow}
\usepackage{booktabs}
\usepackage{color}
\usepackage{tabularx}
\usepackage{setspace}
\usepackage{verbatim}
\usepackage{stfloats}
\usepackage{caption}
\usepackage{float}
\usepackage{url}

\usepackage{stfloats}

\usepackage{amsmath}
\allowdisplaybreaks[0]

\makeatletter
\newif\if@restonecol
\makeatother

\usepackage[linesnumbered,ruled,vlined]{algorithm2e}
\usepackage{algpseudocode}
\usepackage{amsmath}

\usepackage{fancyhdr}

\begin{document}
\begin{spacing}{2.0}

\title{Delay Sensitive Task Offloading in the 802.11p Based Vehicular Fog Computing Systems}

\author{Qiong Wu,~\IEEEmembership{Member,~IEEE}, Hanxu Liu, Ruhai Wang,~\IEEEmembership{Senior Member,~IEEE}, Pingyi Fan,~\IEEEmembership{Senior Member,~IEEE}, Qiang Fan and Zhengquan Li,~\IEEEmembership{Member,~IEEE}

}



\maketitle

\begin{abstract}
 Vehicular fog computing (VFC) is envisioned as a promising solution to process the explosive tasks in autonomous vehicular networks. In the VFC system, task offloading is the key technique to process the computation-intensive tasks efficiently. In the task offloading, the task is transmitted to the VFC system according to the 802.11p standard and processed by the computation resources in the VFC system. The delay of task offloading, consisting of the transmission delay and computing delay, is extremely critical especially for some delay sensitive applications. Furthermore, the long-term reward of the system (i.e., jointly considers the transmission delay, computing delay, available resources and diversity of vehicles and tasks) becomes a significantly important issue for providers. Thus, in this paper, we propose an optimal task offloading scheme to maximize the long-term reward of the system where 802.11p is employed as the transmission protocol for the communications between vehicles. Specifically, a task offloading problem based on a semi Markov decision process (SMDP) is formulated. To solve this problem, we utilize an iterative algorithm based on Bellman equation to approach the desired solution. The performance of the proposed scheme has been demonstrated by extensive numerical results.
\end{abstract}


\IEEEpeerreviewmaketitle

\section{Introduction}
\label{sec1}
Autonomous vehicles are envisioned as a promising technology to ensure road safety and traffic efficiency, which have brought a lot of convenience to people's life in different application scenarios, e.g., traffic accident avoidance and smart parking. In autonomous vehicles, a large number of diverse on-board sensors are employed to collect the ambient information that can be used to monitor the environment around them. As at least $200$ different sensors are expected to be equipped to autonomous vehicle systems \cite{200sensors}, a huge volume of sensed data is generated when autonomous vehicles are on the road \cite{Wujun2017}. It is predicted that autonomous vehicle systems will generate multiple giga bytes (GB) of data per second, typically from equipped high-quality cameras, LiDARs and radars \cite{Cheng2018}\cite{Bisio2019}. The amount of data generated by autonomous vehicle systems will reach around $4000$ GB everyday in 2020 \cite{4000G}. In general, the sensed data are redundant. It needs to further analyze these redundant data to extract useful information, and to make correct reactions accordingly \cite{Cao2019} \cite{Lihe2018}. In other words, autonomous vehicle has intensive computing tasks to process \cite{Tao2017}\cite{HeLi2019}. Faced with such tremendous data processing pressure, how to efficiently process the intensive computation tasks to extract useful information becomes a critical issue.

\begin{figure*}
\centering
\includegraphics[width=5in]{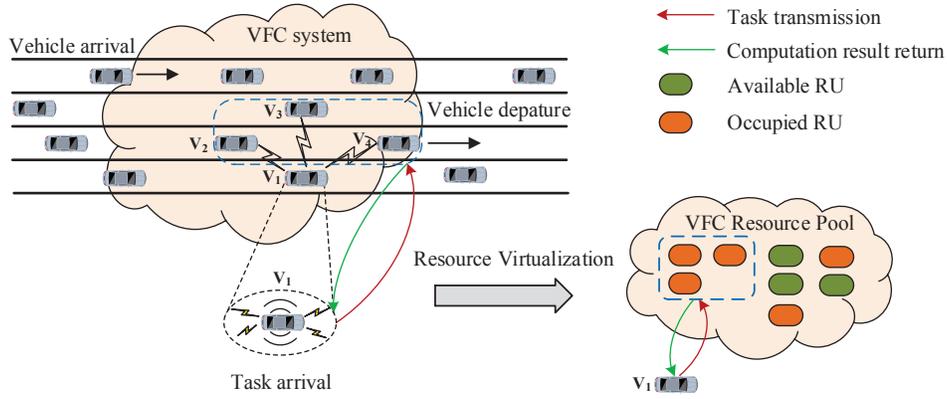}
\caption{The task offloading model in VFC system.}
\label{fig1}
\end{figure*}

With the development of industry, the powerful storage and computing resources of vehicles facilitate the effective task processing on them. In this case, service providers desire to employ the vehicular fog computing (VFC) system (i.e., composed by several vehicles with abundant computing resources) to process the computing tasks \cite{Yao2018}-\cite{Taoming2017}. Since vehicles arrive at or depart from the VFC system randomly, the computing resources in the VFC system change frequently.  In the VFC system, each vehicle has a common resource unit (RU) by leveraging the virtualization technology \cite{Chen2017}\cite{Lin2018} and is aware of the available RUs of the whole system in real time through the communications with other vehicles. All vehicles employ the IEEE 802.11p as the communications standard that has been widely used to provide the reliable and efficient communications in the vehicular environment \cite{802.11p_1}-\cite{Zheng2016}. In the VFC system, each vehicle can offload its tasks to other vehicles. Specifically, as shown in Fig. \ref{fig1}, the system first allocates a certain quantities of RUs to process the tasks of a vehicle (i.e., a request vehicle). Afterwards, the vehicle divides the task into sub-tasks with the equal size according to the number of the allocated RUs and transmits the sub-tasks to vehicles with the allocated RUs based on 802.11p \cite{sub-task}.  These allocated RUs form a group to process the sub-tasks cooperatively. After that, the computing result is transmitted back to the request vehicle.

For autonomous vehicles, the task delay is a critical metric that may determine if the vehicles can react to different situations in real-time to ensure the safety of pedestrians and passengers \cite{WuYuan2018}. Generally, an autonomous vehicle should obtain the computing result within $100ms$ \cite{100ms_1}\cite{100ms_2}, i.e., the delay of the task offloading should not exceed $100ms$. In fact, the task delay consists of the computing delay and transmission delay. The computing delay is the time duration for computing the offloaded task, while the transmission delay stands for the sum time duration of transmitting the task and feedback the computation result. Note that to avoid the collisions between the transmitting tasks and feedback computing results, the vehicles in the VFC systems usually use different transceivers with separate radio channels to transmit tasks and computing results. Similar to the related work \cite{neglect_delay1}\cite{neglect_delay2}, since the computing result has a much small size, the delay incurred by transmitting the computation result is much smaller as compared to that of tasks and is neglected here. In this paper, the transmission delay is referred to the delay for transmitting the task.

In the task offloading, both the computing delay and transmission delay are affected by the number of the allocated RUs. On one hand, the computing capacity of the group of the allocated RUs increases with the increasing number of allocated RUs, and thus improve the computing delay accordingly. On the other hand, for the transmission delay, the request vehicle in the VFC system adopts the 802.11p standard to transmit the sub-tasks to each of the allocated RUs in turn. Therefore, the number of transmissions is increased with the number of the increasing allocated RUs, and thus deteriorates the transmission delay. Moreover, the computing resources are limited and change frequently in the VFC system owing to the random arrival and departure of vehicles in the system. It may affect the available resources of the VFC system. Considering the above factors, the providers need to improve the long-term reward of the system in the task offload,  where the transmission delay, computing delay, available RUs and diversity feature of tasks and vehicles are taken into account (i.e., including both the gain and cost of the system). To solve the problem, a proper offloading scheme has to be designed to yield the suitable decision on where to offload each vehicle's tasks, and thus maximize the long-term reward of the system. To the best of our knowledge, no work has considered the transmission delay caused by the 802.11p standard in the task offloading to maximize the long-term rewards.

In this paper, we propose an optimal offloading scheme to maximize the long-term rewards of the system where 802.11p is employed as the transmission protocol for the communications between vehicles. The main contributions of our work are summarized as follows.

\begin{itemize}
\item[1)] We jointly consider the transmission delay, computing delay, available RUs and the diversity feature of vehicles and tasks in the task offloading. Specifically, we design a discounted semi-Markov decision process (SMDP) to formulate the task offloading problem in the VFC system. In  the SMDP model,  SMDP states, actions, rewards and transition probabilities are formulated for analytical tractability. To solve the problem, we utilize an iterative algorithm to achieve the maximal long-term reward of the system.

\item[2)] Given 802.11p, we derive and analyze the transmission delay and task arrival rate under different decisions.   Moreover, we analyze the maximum number of vehicles in the VFC system to ensure that the task delay meets the maximum delay limit.

\item[3)]The performance of the proposed scheme has been demonstrated by extensive numerical results. As compared with baseline algorithms, the proposed algorithm can significantly improve the long-term reward.

\end{itemize}

The rest of the paper is organized as follows. Section \ref{sec2} provides a review of related work. Section \ref{sec3} describes the VFC system model. The SMDP that formulates the offloading problem is described in Section \ref{sec4}. The iteration algorithm to obtain the optimal task offloading scheme is given in Section \ref{sec5}. The numerical results are shown in Section \ref{sec6}. Section \ref{sec7} concludes this paper and also lists the future work directions.

\section{Related Work}
\label{sec2}
 In recent years, researchers carried out a few of studies on task offloading in VFC.
 In \cite{Hou2016}, Hou \emph{et al.} conceived the concept of a VFC system including moving and parked vehicles to compute task collaboratively. They conducted a quantitative analysis to analyze the capacities of the VFC system in four typical scenarios and draw the conclusion that the computational performance of the VFC system can be improved greatly compared with the cloud computing.
 In \cite{Ning2019}, Ning \emph{et al.} presented a three-layer VFC system architecture for the real-time citywide traffic management in smart cities. They further proposed an offloading scheme to minimize the response time of the citywide events collected by vehicles and tested the performance by using the real-world taxi trajectory.
 In \cite{Wang2019}, Wang \emph{et al.} considered the system cost including the service delay and energy consumption as a new service arrives in the VFC system, and proposed a dynamic reinforcement learning scheduling algorithm based on the Markov decision process to obtain the offloading scheme in order to minimize the system cost.
In \cite{ZhangYi2019}, Zhang \emph{et al.} proposed an auction scheme which is used to guide the moving vehicles to the available parking places with less cost in the VFC system, where the parked vehicles contribute their computation capabilities to compute the delay-sensitive services of the moving vehicles.
In \cite{Liu2018}, Liu \emph{et al.} presented a fog and cloud integrated computing system with three-layer architecture where the tasks can be offloaded to the nearby fog nodes or cloud center. They jointly considered the constrained computation, storage and spectrum resources to design the offloading scheme in order to minimize the system cost.
In \cite{Zhou2019}, Zhou \emph{et al.} studied the computation resource allocation and task assignment problem in the VFC system respectively, and proposed an incentive mechanism based on the contract theory to encourage the near vehicles share their computation resources, thus the computation tasks can be offloaded from the base station to the shared computation resources.
In \cite{WangZhe2018}, Wang \emph{et al.} considered the heterogeneous delay requirements as well as the dynamic topology of the VFC system and proposed an application-aware offloading scheme to maximize the long-term reward of the system.
In \cite{ZhangWen2017}, Zhang \emph{et al.} proposed a regional cooperative VFC system to deal with a mass of data in the smart city and presented a resource management scheme to optimize packet dropping rates and energy efficiency in a hierarchical model consisted of inter-fog and intra-fog network.
In \cite{Zhu2019}, Zhu \emph{et al.} considered the service latency, quality loss and fog node capacity in the VFC system, and formulated the task allocation process as a bi-objective optimization problem. The optimal task allocation approach is obtained by a binary particle swarm optimization algorithm.

From the mentioned above, the task offloading of the VFC system have been studied by some literatures. However, there is no literature on maximizing the long-term reward where the transmission delay caused by the 802.11p standard, computing delay, available RUs and the variability feature of vehicles and tasks are jointly taken into consideration, which motivated us to conduct this work.

\section{System Model}
\label{sec3}
 In this section, we describe the system model in detail. The scenario is first introduced and then the procedure of transmitting a sub-task according to the 802.11p standard is described.

 In the scenario considered in this paper, vehicles moving on a highway with multiple lanes form a dynamic VFC system to process the tasks. The vehicles join or leave the VFC system according to the Poisson process with rate $\lambda_f$ and $\mu_f$, respectively \cite{Sun2018}\cite{Zheng2015}. In the VFC system, the vehicles adopt the 802.11p standard to communicate with each other through one-hop communication. In order to find the system capability limit, we consider its extreme scenario, the system is saturated, i.e., each vehicle in the VFC system always has a task to offload. In general, when the system is not saturated, it will use less transmission delay and computing delay. Note that each vehicle has the same computing resource, i.e., a RU. As service rate of a RU is $\mu_t$, the service rate of $i$ RUs, is $i\mu_t$. Each vehicle knows the available RUs in the whole system by communicating with each other. When the request vehicle has a task to offload, the system makes a decision to allocate RUs to this task. Afterwards, the request vehicle divides the task into the sub-tasks with the same size according to the number of allocated RUs and then adopts the 802.11p standard to transmit the sub-tasks to their destination RUs. The detailed procedure of transmitting a sub-task is introduced in the next paragraph. Afterwards, these allocated RUs form a group to process the task cooperatively and then feedback to the request vehicle. The scenario considered in this paper is shown in Fig. \ref{fig1}.

\begin{figure*}
\centering
\includegraphics[width=5in]{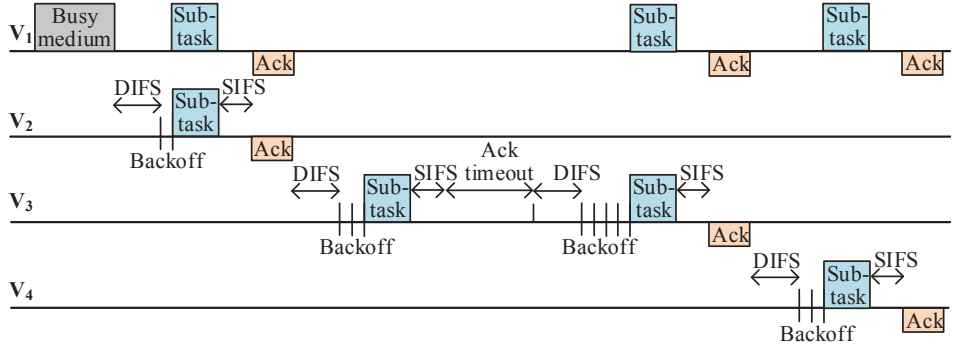}
\caption{The process of transmitting a task to the allocated RUs.}
\label{fig2}
\end{figure*}

 The procedure of transmitting a sub-task is introduced as follows. The IEEE 802.11p enhanced distribute channel function (DCF) mechanism is adopted to transmit each packet \cite{802.11p_1}. Specifically, when a vehicle has a packet to transmit, it will initialize a back-off counter with a value selected randomly from $0$ to $W-1$, where $W$ is the contention window and equals to the minimum contention window $W_{min}$. Then the value of the back-off counter is reduced by $1$ after each slot. Once the back-off counter becomes $0$ the packet is transmitted. After the packet is transmitted, if an acknowledge (ACK) message is received after a short inter-frame space (SIFS) interval, the transmission is considered to be successful. If an ACK message is not received within an ACK timeout interval, the transmission is unsuccessful and the packet need to be retransmitted. Before the retransmission, the contention window $W$ is doubled, i.e., $W=2W_{min}$. Then a new back-off procedure restarts where the back-off counter is initialized and reconfigured from $0$ to $2W_{min}-1$. If the contention window is doubled for $m$ times, the contention window would keep the value $2^mW$. The above retransmission repeats until the packet is successfully delivered. After the packet is transmitted successfully, the contention window is reset to $W_{min}$ and a new back-off procedure is initialized after a distributed inter-frame space (DIFS) for another transmission.

The task is transmitted successfully when all the sub-tasks are delivered to the allocated RUs in turn. The process that vehicle $V_1$ adopts the 802.11p standard to transmit a task to the $3$ RUs $V_2$, $V_3$ and $V_4$ in Fig. \ref{fig1} is illustrated in Fig. \ref{fig2}.

\begin{table}
\footnotesize
\caption{The notations used in this paper.}
\label{tab1}
\centering
\begin{tabular}{|c|p{6.6cm}|}
\hline
\textbf{Notation} &\textbf{Description}\\
\hline
$K$ &Maximum number of vehicles in the VFC system.\\
\hline
$M$ &Number of RUs in the VFC system.\\
\hline
$n_i$ &Number of the tasks that are processed by $i$ RUs.\\
\hline
$N$ &\multicolumn{1}{m{6.6cm}|}{Maximal number of RUs that the VFC system can allocate.}\\
\hline
$\lambda_{f}/\mu_{f}$ &Arrival / departure rate of vehicles.\\
\hline
$\lambda_{t}(i)$ &\multicolumn{1}{m{6.6cm}|}{Arrival rate of tasks when the current task is offloaded to $i$ RUs.}\\
\hline
$\bar\lambda_{t}$ &Expected task arrival rate under different decisions.\\
\hline
$p_{i}$ &The probability that a task is allocated to $i$ RUs.\\
\hline
$\mu_{t}$ &Computation service rate of a RU.\\
\hline
$A$ &Arrival of a task.\\
\hline
$D_{i}$ &Departure of a task which is processed by $i$ RUs.\\
\hline
$F_{+1}/F_{-1}$ &Arrival / departure of a vehicle.\\
\hline
$W$ &Contention window.\\
\hline
$W_{min}$ &Minimum contention window.\\
\hline
$m$ &Maximum back-off stage.\\
\hline
$p$ &Collision probability.\\
\hline
$\tau$ &Transmission probability.\\
\hline
$E[D_i]$ &\multicolumn{1}{m{6.6cm}|}{Average delay of transmitting a subtask to one of the allocated $i$ RUs.}\\
\hline
$E[N]$ &\multicolumn{1}{m{6.6cm}|}{Average number of slots required for transmitting a sub-task successfully.}\\
\hline
$T_{slot}(i)$ &\multicolumn{1}{m{6.6cm}|}{Average time occupied by a slot when the sub-task is transmitted to one of the allocated $i$ RUs.}\\
\hline
$P_{idle}$ &The probability that an idle slot is detected.\\
\hline
$P_{c}$ &The probability that a collision occurs.\\
\hline
$P_{s}$ &The probability that a successful transmission occurs.\\
\hline
$slot$ &The duration time of an idle slot.\\
\hline
$T_{c}(i)$ &\multicolumn{1}{m{6.6cm}|}{The duration time of a collision when the sub-task is transmitted to one of the allocated $i$ RUs.}\\
\hline
$T_{s}(i)$ &\multicolumn{1}{m{6.6cm}|}{The duration time of a successful transmission the sub-task is transmitted to one of the allocated $i$ RUs.}\\
\hline
$D_{t}(i)$ &\multicolumn{1}{m{6.6cm}|}{The transmission delay that the request vehicle transmits the task to $i$ RUs.}\\
\hline
$D_{p}(i)$ &The computing delay that the task is processed by $i$ RUs.\\
\hline
$\beta$ &The saved price per unit time.\\
\hline
$T$ &The required time to process the task locally.\\
\hline
$\xi$ &The punishment that the VFC system drops the task.\\
\hline
$\eta$ &\multicolumn{1}{m{6.6cm}|}{The punishment that a vehicle departs when there is no available RUs in the VFC system.}\\
\hline
$\alpha$ &Continuous-time discount factor.\\
\hline
\end{tabular}
\end{table}

\section{Problem Formulation}
\label{sec4}

In the VFC system, the total number of available RUs is affected by the events, such as the arrival and departure of a vehicle, and the arrival and departure of a task. Each vehicle gets the updated information on available RUs in real time. In these events, when a task from a request vehicle arrives at the VFC system, the system makes a decision to allocate a certain number of RUs to process this task, while the VFC system does not need to make any decision for the other events. By making a task offloading decision, the system achieves a reward which depends on the transmission delay, computing delay, current available RUs and the diversity feature of vehicles and tasks.

In this section, we employ an SMDP model to formulate this problem. In the SMDP model, the states stand for a set that consists of the number of the tasks allocated with different number of RUs and the number of the available RUs under different events; the actions reflect the quantities of allocated RUs under different events; the rewards reflect the benefit of the VFC system under different states and actions; the transition probabilities reflect the probabilities of the state transition under different actions. Next, the states, actions, rewards and transition probabilities are formulated respectively. The main notations used in this paper are summarized in Table \ref{tab1}.

\subsection{States}
The states in the SMDP are formulated to indicate the processing tasks with different number of RUs and the number of the available RUs under different events. The event is denoted by $e$, where  $e$ is a specific event which belongs to the set $\left\{A, D_{1}, \ldots, D_{N}, F_{+1}, F_{-1} \right\}$. Here, we denote $A$ as the arrival of a task, $N$ as the maximal number of RUs that the system can allocate, $D_{i}$ as the departure of the task which is processed by $i$ RUs$(1\leq i\leq N)$, $F_{+1}$ as the arrival of a vehicle, and $F_{-1}$ as the departure of a vehicle. Furthermore, state with event $e$ is denoted by $s=\left(M, n_{1}, \ldots, n_{N}, e\right) $, where $M$ is the number of RUs in the VFC system; $n_{i}$ is the number of tasks that are processed by $i$ RUs. Thus, the set of the states can be denoted as

 \begin{equation}
S=\left\{s | s=\left(M, n_{1}, \ldots, n_{N}, e\right)\right\}.
\label{eq1}
\end{equation}

In the system, the number of RUs allocated to tasks cannot exceed the total number of RUs, i.e., $\sum_{i=1}^{N} i \cdot n_{i} \leqslant M$ and the number of RUs $M$ cannot exceed $K$ (i.e., the maximal number of vehicles in the VFC system). In addition, the number of available RUs can be calculated as $M-\sum_{i=1}^{N} i \cdot n_{i}$.

\subsection{Actions}
The actions in the SMDP are formulated to indicate the decisions to allocate a certain number of RUs under different events. The action based on the state $s$ is denoted by $a(s)$ and belongs to the set $\{-1,0,1,2, \ldots, N\}$. Here, $a(s)=-1$ indicates that no action is taken; $a(s)=0$ indicates that the VFC system rejects to offload a task among vehicles and drops the task when the available RUs are insufficient; $a(s)=i$ means that $i$ RUs are allocated to process the task. When an event (such as the departure of a task, the departure of a vehicle and the arrival of a vehicle) happens, the VFC system may have no task to process and takes no action. When the event such as the arrival of a task happens, the VFC system may drop the task or allocate $i$ RUs to process the task. The relationship between the events and the actions are shown in Eq. (\ref{eq2}).
\begin{equation}
A_{c}=\left\{
\begin{array}{ll}{\{-1\},} & {e \in\left\{D_{1}, \ldots, D_{N}, F_{+1}, F_{-1} \right\}} \\
{\{0,1,2, \ldots, N\}}, & {e=A}
\end{array}\right.
\label{eq2}
\end{equation}

\subsection{Rewards}
The rewards are formulated as the benefit of the VFC system when different actions are taken at different states, where the transmission delay, computing delay, current available resources and the diversity feature of vehicles and tasks are jointly taken into account.  When action $a$ is taken at state $s$, the system obtains an immediate income $I(s, a)$. The state $s$ will be hold for a time duration until the state $s$ is transitioned to the next state when another event occurs. In this duration, the cost of the system is denoted by $C(s, a)$. When action $a$ is taken at state $s$, the reward $R(s, a)$ is the difference between the income $I(s, a)$ and cost $C(s, a)$, and thus can be expressed
\begin{equation}
\begin{array}{l}
R(s, a)=I(s, a)-C(s, a).
\label{eq3}
\end{array}
\end{equation}

We will further describe the income $I(s, a)$ and cost $C(s, a)$ as follows:

\subsubsection{Income}
 As states change based on the events, the incomes are different under different actions and events. Thus, we will formulate the incomes under different actions and events, respectively.

\begin{itemize}
\item[(a)] $a=i, e=A$\\
When a task arrives (i.e., event $A$) occurs and the available RUs are sufficient, the VFC system allocates $i$ RUs to process the task. In this case, the task offloading can reduce the task delay as compared with processing the task locally. As task delay is extremely sensitive in the VFC system, the savement of task delay is formulated as the income of the system. Since the task delay consists of the computing delay and the transmission delay, the immediate income is denote as $\beta \cdot[T-D_{t}(i)-D_{p}(i)]$, where $\beta$ is the saved price per unit time; $T$ is the computing delay when a task is processed locally, $D_{t}(i)$ is the transmission delay that the request vehicle transmits the task to $i$ RUs; $D_{p}(i)$ is the computing delay that the task is processed by $i$ RUs on different vehicles.

\item[(b)] $a=0, e=A$\\
When a task arrives and the available RUs are insufficient, the VFC system rejects to offload the task and then drops it. In this case, the request vehicle cannot obtain the result in the task and thus cannot make a reaction accordingly. Therefore, this decision is very detrimental to system, thus the VFC system is punished with $\xi$ and the SMDP is terminated.

\item[(c)] $a=-1, e=\in\left\{D_{1}, \ldots, D_{N}, F_{+1}\right\}$\\
When the event such as the departure of a task or arrival of a vehicle occurs, the VFC system takes no action. In this case, the VFC system has no income.

\item[(d)] $a=-1, e=F_{-1}, \sum_{j=1}^{N} {n}_{j} j \textless M$\\
When the event that a vehicle departs the system occurs, the VFC system takes no action. In this case, if there are available RUs in the VFC system, the VFC system has no income.

\item[(e)] $a=-1, e=F_{-1}, \sum_{j=1}^{N} {n}_{j} j = M$\\
When the departure of a vehicle occurs, the VFC system takes no action. If there are no available RUs in the VFC system, i.e., all vehicles in the VFC system are processing a task, the departure of a vehicle would interrupt the processing of a task. In this case, the VFC system is punished with $\eta$ and the SMDP is terminated.
\end{itemize}

Conclusively, the incomes under different actions and events are derived as:
\begin{spacing}{1.35}
\begin{equation}
I(s,a)=\\
\left\{
\begin{array}{l}
{\beta \cdot[T-D_{t}(i)-D_{p}(i)]},\\
\qquad\quad\ {a=i, e=A(i>0)}\\
{-\xi, \qquad a=0, e=A}\\
{0, \qquad\quad a=-1, e \in\left\{D_{1}, \ldots, D_{N}, F_{+1}\right\}} \\
{0, \qquad\quad  a=-1, e=F_{-1}, \sum_{j=1}^{N} {n}_{j} j \textless M}\\
{-\eta,} \quad\quad\  {a=-1, e=F_{-1}, \sum_{j=1}^{N} {n}_{j} j=M}\\
\end{array}
\right.
\label{eq4}
\end{equation}
\end{spacing}

 Here, the computing delay $D_{p}(i)$ is the time for processing the task by $i$ RUs, which depends on the total computing capacity of the allocated RUs. In this paper, we assume each RU has the same computing capacity. Here, we use the computing service rate to reflect the computing capacity. The service rate of each RU is denoted as $u_t$, thus the service time to process the task by $i$ RUs is calculated as
\begin{equation}
D_{p}(i)=\frac{1}{i\cdot u_t}.
\label{eq5}
\end{equation}

As the task is divided into sub-tasks with equal sizes based on the number of allocated RUs $i$, the sub-tasks are transmitted to the allocated $i$ RUs. Therefore, the delay of transmitting the task to  $i$ RUs is
\begin{equation}
D_t(i)=i \cdot E[D_i],
\label{eq6}
\end{equation}
where $E[D_i]$ is the average delay of transmitting a sub-task to one of the allocated $i$ RUs. According to the 802.11p DCF mechanism, the value of the backoff counter is decreased by one after each slot. The status of a slot may be successful transmission, collision or idle, thus the average delay of transmitting a packet is calculated as
\begin{equation}
E[D_i]=E[N] \cdot T_{slot}(i),
\label{eq7}
\end{equation}
 where $E[N]$ is the average number of slots required for transmitting the sub-task successfully; $T_{slot}(i)$ is the average slot time.

Considering the probabilities of successful transmission, collision and idle in a slot, $T_{slot}(i)$ is calculated as

\begin{equation}
T_{slot}(i)=P_{idle} \cdot slottime+P_{c} \cdot T_{c}(i)+P_{s} \cdot T_{s}(i),
\label{eq8}
\end{equation}
where $P_{s}$, $P_{c}$ and $P_{idle}$ are the probabilities of successful transmission, collision and idle, respectively.  $T_{s}(i)$, $T_{c}(i)$ and $slottime$ are the time duration of a slot at the statuses of  successful transmission, collision and idle, respectively.

Specifically, $T_{c}(i)$ and $T_{s}(i)$ are calculated as follows \cite{Tc_Ts_1},
\begin{equation}
T_{s}(i)=H+E[P]/i+\mathrm{SIFS}+\delta+\mathrm{ACK}+\delta+\mathrm{DIFS},
\label{eq9}
\end{equation}
\begin{equation}
T_{c}(i)=H+E[P]/i+\mathrm{SIFS}+\delta+\mathrm{ACK timeout},
\label{eq10}
\end{equation}
where $H$ is the length of the packet header and $E[P]$ is the length of the task. Since the task is divided into $i$ sub-tasks with the same sizes, the sub-task length is $E[P]/i$. $\mathrm{SIFS}$,  $\mathrm{DIFS}$, $\mathrm{ACK}$ are the length of the SIFS, DIFS and ACK (i.e, control packets), respectively; $\mathrm{ACK timeout}$ is the length of the ACK time out interval and can be expressed as $\mathrm{ACK timeout}= \mathrm{ACK}+\delta+\mathrm{DIFS}$ where $\delta$ denote the propagation delay of a packet.

Next, $P_{s}$, $P_{c}$ and $P_{idle}$ in Eq. (\ref{eq8}) are derived. We denote $p$ as the probability of collision, where more than one vehicle are transmitting data simultaneously, and $\tau$ as of transmission probability that a vehicle is transmitting data. As we know, in a slot at the status of successful transmission, only one vehicle is transmitting; for the collision status, more than one vehicle are transmitting; for the idle status, no vehicle is transmitting in the system. Therefore, $P_{idle}$, $P_{s}$ and $P_{c}$ in Eq. (\ref{eq8}) are calculated as follows,
\begin{equation}
\begin{array}{l}
P_{i d l e}=(1-\tau)^{M},
\end{array}
\label{eq11}
\end{equation}
\begin{equation}
\begin{array}{l}
P_{s}=M \tau(1-\tau)^{M-1},
\end{array}
\label{eq12}
\end{equation}
\begin{equation}
\begin{array}{l}
P_{c}=1-P_{idle}-P_{s},
\end{array}
\label{eq13}
\end{equation}
where $\tau$ and $p$ can be calculated as Eq. (\ref{eq14}) and (\ref{eq15}) according to \cite{bianchi2000},
\begin{equation}
\tau=\frac{2(1-2p)}{(1-2 p)(W+1)+pW\left(1-(2p)^{m}\right)},
\label{eq14}
\end{equation}
\begin{equation}
\begin{array}{l}
p=1-(1-\tau)^{M-1}.
\label{eq15}
\end{array}
\end{equation}

Based on Eq. (7), in order to calculate the average delay of transmitting a packet, the average number of slots ($E[N]$) need to be determined first.

Since the contention window is doubled in each retransmission when the number of retransmissions reaches $m$, $E[N]$ consists of two pars, i.e.,
\begin{equation}
\begin{array}{c}
E[N]=E[N_1]+E[N_2],
\label{eq16}
\end{array}
\end{equation}

where $E[N_1]$ and $E[N_2]$ are the average number of slots when the number of retransmissions is not larger than $m$ and larger than $m$, respectively. At first, $h$ is denoted as the required number of retransmissions before successfully delivering a packet, i.e., the transmission repeats for $h+1$ times. Therefore, the probability that the transmission is successful is $p^{h}(1-p)$. Next, for a given retransmission, we will further derive the average required number of slots. Let the contention window be $W_l$ when the  number of retransmissions is $l$. Since a value of the backoff counter is first selected from [0, $W_l$-1] randomly with probability  $\frac{1}{W_l}$ and then is decreased by one after each slot until the value is decreased to $0$ in the procedure of the retransmission, the average number of slots for the $l$th retransmission is calculated as $\sum_{k=1}^{W_{l}} \frac{k}{W_{l}}= \frac{W_{l}+1}{2}$. Note that, the contention window keeps a maximum contention $W_m$ when $m\leq l$. Therefore, $E[N_1]$ and $E[N_2]$ can be calculated as Eq. (\ref{eq17}) and Eq. (\ref{eq18}), respectively. Eq. (\ref{eq17}) and Eq. (\ref{eq18}) are shown at the top of this page. By summing $E[N_1]$ and $E[N_2]$, the average number of slots $E[N]$ is given as Eq. (\ref{eq19}), shown at the top of this page.

\begin{equation}
\begin{aligned}
&E\left[N_{1}\right]=\sum_{h=0}^{m} p^{h}(1-p) \sum_{l=0}^{h} \frac{W_{l}+1}{2}=\frac{1-(m+2) p^{m+1}+(m+1)p^{m+2}}{2(1-p)}+\frac{(1-p)\left[1-(2 p)^{m+1}\right]W}{1-2 p}-\frac{(1-p^{m+1})W}{2}
\end{aligned}
\label{eq17}
\end{equation}

\begin{equation}
\begin{aligned}
&E\left[N_{2}\right]=\sum_{h=m+1}^{+\infty} p^{h}(1-p)\left[\sum_{l=0}^{m} \frac{W_{l}+1}{2}+\frac{W_{m}+1}{2}(h-m+1)\right]=\frac{p^{m+1}}{2}\left[m+1+\left(2^{m+1}-1\right) W+\frac{(2-p)(2^{m} W+1)}{1-p}\right]
\end{aligned}
\label{eq18}
\end{equation}

\begin{equation}
\begin{aligned}
E[N]&=E[N_{1}]+E[N_{2}]\\
&=\frac{1-(m+2) p^{m+1}+(m+1)p^{m+2}}{2(1-p)}+\frac{(1-p)\left[1-(2 p)^{m+1}\right]W}{1-2 p}-\frac{(1-p^{m+1})W}{2}\\
&+\frac{p^{m+1}}{2}\left[m+1+\left(2^{m+1}-1\right) W+\frac{(2-p)(2^{m} W+1)}{1-p}\right]
\end{aligned}
\label{eq19}
\end{equation}

As a result, the average time of a slot and the average number of slots for delivering a packet can be calculated according to  Eq. (\ref{eq8}) and Eq. (\ref{eq19}) respectively. Thus, the average delay of transmitting a packet can be calculated according to Eq. (\ref{eq7}).

\subsubsection{Cost}
The discounted cost model is adopted to formulate the long-term cost. $C(s,a)$ is the expected discounted cost of the VFC system during the time duration between taking action $a$ and the corresponding state transition. Similar with \cite{Zheng2015}, the duration is assumed to follow an exponentially distribution. According to \cite{Puterman1970} and \cite{Puterman2005}, the expected discounted cost is expressed by
\setcounter{equation}{19}
\begin{equation}
\begin{aligned}  C(s, a)&=b(s, a) E_{s}^{a}\left\{\int_{0}^{\tau} e^{-\alpha t} d t\right\} \\ &=b(s, a) E_{s}^{a}\left\{\frac{1-e^{-\alpha \tau}}{\alpha}\right\} \\& =\frac{b(s, a)}{\alpha+\sigma(s, a)} \end{aligned},
\label{eq20}
\end{equation}
where $\alpha$ is the discount factor, $b(s,a)$ is the cost rate of the expected service time under state $s$ and action $a$ which can be expressed as a function of the number of allocated RUs, i.e.,

 \begin{equation}
 b(s, a)=\sum_{i=1}^{N} i \cdot n_{i}.
\label{eq21}
\end{equation}
$\sigma(s, a)$ is the expected event rate under state $s$ and action $a$. It denotes the sum of the arrival and departure rate of all events in the VFC system under state $s$ and action $a$. The arrival rate and departure rate of vehicles are $\lambda_f$ and $\mu_f$, respectively. We have analyzed the arrival rate and departure rate of tasks under different events and actions as follows:

\begin{itemize}

\item[(a)] $a=i, e=A$\\

When a task arrives at the VFC system occurs and the system allocates $i$ RUs to process the task, the task arrival rate in the system is $M\lambda_{t}(i)$, where $\lambda_{t}(i)$ is the task arrival rate under action $i$ and can be calculated by $\frac{1}{E[D_i]}$. Meanwhile, departure rate of tasks is calculated by $\left(\sum_{j=1}^{N} {n}_{j} j+i\right) \mu_{t}$ due to the number of allocated RUs under action $i$ is $\left(\sum_{j=1}^{N} {n}_{j} j+i\right)$.\\

\item[(b)] $a=-1, e=D_i$\\

When the event that a task allocated $i$ RUs and departs from the system occurs, the system takes no action. Since the system is saturated, i.e., each vehicle always has a task to transmit, the event happens when at least one request vehicle has been allocated RUs in the previous states and is transmitting the task. Due to the previous states and actions are not recorded in the SMDP, the number of allocated RUs is uncertain, and thus the expected task arrival rate under different actions is adopted to calculate the arrival rate of tasks in this case. Hence, the arrival rate of tasks is $M \bar\lambda_{t}$, where $\bar\lambda_{t}$ is the expected task arrival rate under different actions and is calculated as
\begin{equation}
\bar\lambda_{t}=\sum_{i=1}^{N}p_{i}\lambda_{t}(i),
\label{eq22}
\end{equation}
$p_i$ is the probability that a task is allocated with $i$ RUs when it arrives at the system. It is estimated according to ratio between the number of tasks allocated with $i$ RUs and the total number of the tasks, i.e.,
\begin{equation}
p_{i}=\frac{n_{i}}{\sum_{i=j}^{N}n_{j}}.
\label{eq23}
\end{equation}

The departure rate of the tasks is $\left(\sum_{j=1}^{N} {n}_{j} j-i\right) \mu_{t}$.\\

\item[(c)] $a=-1, e=F_{+1}$\\

 When the event that a vehicle arrives at the system occurs, the system takes no action. In this case, the number of vehicles will be increased by $1$, thus the arrival rate of tasks is $(M+1) \bar\lambda_{t}$. Meanwhile, the departure rate of vehicles is $\sum_{j=1}^{N} {n}_{j} j\mu_{t}$.\\

\item[(d)] $a=-1, e=F_{-1}$\\

 When the event that a vehicle departs from the system occurs, the system takes no action. In this case, the number of vehicles will be decrease by $1$, thus the arrival of the tasks is $(M-1) \bar\lambda_{t}$. The departure rate of tasks is $\sum_{j=1}^{N} {n}_{j} j\mu_{t}$.

\end{itemize}

\begin{equation}
\sigma(s, a)=\frac{1}{\tau(s, a)}=\left\{
\begin{array}{ll}
M \lambda_{t}(i)+\lambda_{f}+\mu_{f}+\left(\sum_{j=1}^{N} {n}_{j} j+i\right) \mu_{t}, \quad e=A, a=i (1 \leq i \leq N)\\
M \bar\lambda_{t}+\lambda_{f}+\mu_{f}+\left(\sum_{j=1}^{N} {n}_{j} j-i\right) \mu_{t}, \qquad e=D_{i}, a=-1 \\
(M+1) \bar\lambda_{t}+\lambda_{f}+\mu_{f}+\sum_{j=1}^{N} {n}_{j} j \mu_{t}, \qquad\ e=F_{+1}, a=-1\\
(M-1) \bar\lambda_{t}+\lambda_{f}+\mu_{f}+\sum_{j=1}^{N} n_{j} j \mu_{t}, \qquad\ e=F_{-1}, a=-1
\end{array}
\right.
\label{eq24}
\end{equation}

Conclusively, the expected service rate under different actions and events is calculated by Eq. (\ref{eq24}), shown at the bottom of this page. Finally, the  expected cost of the VFC system is obtained through substituting Eq. (\ref{eq21}) and  Eq. (\ref{eq24}) into Eq. (\ref{eq20}).

\subsection{Transition Probabilities}
 In the SMDP, the next state depends on the current state and action, and thus we formulate the transition probability as the probability of reaching the next state after taking an action in the current state. The transition probability from state $s$ to state $s'$ with action $a$ is denoted by $P(s'|s,a)$, where state $s =\left(M, n_{1}, \ldots, n_{N}, e\right)$ and $e$ is the current event. According to the types of the current event and action, the definition of $P(s'|s,a)$ is divided into four cases.

The first case is that when a task arrives, the system takes an action to allocate $i$ RUs to the task. As the next event is the arrival of a task, the transition probability can be expressed as the ratio between the task arrival rate and the sum of the rate of all events, i.e., $\frac{M \lambda_{t}(i)}{\sigma(s, a)}$. If the next event is the departure of a task required to be processed by $i$ RUs, the value of $n_i$ in state $s$ is increased by $1$, and thus the transition probability is the ratio between the departure rate of the $n_i+1$ tasks processed by $i$ RUs and the sum of the arrival and departure rate of all events, i.e., $\frac{\left(n_{i}+1\right) i \mu_{t}}{\sigma(s, a)}$. If the next event is the departure of a task processed by $j$ RUs ($i \neq j$), the transition probability is the ratio between the departure rate of the $n_j$ tasks processed by $j$ RUs and the sum of the arrival and departure rate of all events, i.e., $\frac{n_{j} j \mu_{t}}{\sigma(s, a)}$. If the next event is the arrival or departure of a vehicle, the transition probability is the ratio between the arrival rate or departure rate of vehicles and sum of the arrival and departure rate of all events, respectively, i.e., $\frac{\lambda_{f}}{\sigma(s, a)}$ and $\frac{\mu_{f}}{\sigma(s, a)}$. Therefore, given the current event $A$ and action $i$, the transition probability is formulated as Eq. (\ref{eq25}).

\subsubsection{$s=\left(M, n_{1}, \ldots, n_{N}, A\right), a=i$}
\setcounter{equation}{24}
\begin{equation}
\begin{split}
P(s'|s,a)=\qquad\qquad\qquad\qquad\qquad\qquad\qquad\qquad\qquad\qquad\\
\left\{
\begin{array}{l}
{\frac{M \lambda_{t}(i)}{\sigma(s, a)}, \quad s^{\prime}=\left(M, n_{1}, \ldots, n_{i}+1, \ldots, n_{N}, A\right)} \\
{\frac{\left(n_{i}+1\right) i \mu_{t}}{\sigma(s, a)}, s^{\prime}=\left(M, n_{1}, \ldots, n_{i}+1, \ldots, n_{N}, D_{i}\right)} \\
{\frac{n_{j} j \mu_{t}}{\sigma(s, a)}, \quad i \neq j,}\\ \qquad\qquad {s^{\prime}=\left(M, n_{1}, \ldots, n_{i}+1, \ldots, n_{N}, D_{j}\right)}\\
{\frac{\lambda_{f}}{\sigma(s, a)}, \quad s^{\prime}=\left(M, n_{1}, \ldots, n_{i}+1, \ldots, n_{N}, F_{+1}\right)} \\
{\frac{\mu_{f}}{\sigma(s, a)}, \quad s^{\prime}=\left(M, n_{1}, \ldots, n_{i}+1, \ldots, n_{N}, F_{-1}\right)}
\end{array}
\right.
\end{split}
\label{eq25}
\end{equation}

When the current event is $D_i$, $F_{+1}$ and $F_{-1}$, the action is $-1$. Similarly, the transition probabilities in these cases are formulated as Eq. (\ref{eq26}), Eq. (\ref{eq27}) and Eq. (\ref{eq28}), respectively.

\subsubsection{$s=\left(M, n_{1}, \ldots, n_{N}, D_{i}\right), a=-1$}
\begin{equation}
\begin{split}
P(s'|s,a)=\qquad\qquad\qquad\qquad\qquad\qquad\qquad\qquad\qquad\qquad\\
\left\{
\begin{array}{l}
{\frac{M \bar\lambda_{t}}{\sigma(s, a)}, \quad, s^{\prime}=\left(M, n_{1}, \ldots, n_{i}-1, \ldots, n_{N}, A\right)} \\
{\frac{\left(n_{i}-1\right) i \mu_{t}}{\sigma(s, a)}, s^{\prime}=\left(M, n_{1}, \ldots, n_{i}-1, \ldots, n_{N}, D_{i}\right)}\\
{\frac{n_{j} j \mu_{t}}{\sigma(s, a)}, \quad i \neq j,}\\ \qquad\qquad
{s^{\prime}=\left(M, n_{1}, \ldots, n_{i}-1, \ldots, n_{N}, D_{j}\right)}\\
{\frac{\lambda_{f}}{\sigma(s, a)}, \quad s^{\prime}=\left(M, n_{1}, \ldots, n_{i}-1, \ldots, n_{N}, F_{+1}\right)}\\
{\frac{\mu_{f}}{\sigma(s, a)}, \quad s^{\prime}=\left(M, n_{1}, \ldots, n_{i}-1, \ldots, n_{N}, F_{-1}\right)}
\end{array}
\right.
\end{split}
\label{eq26}
\end{equation}

\subsubsection{$s=\left(M, n_{1}, \ldots, n_{N}, F_{+1}\right), a=-1$}

\begin{equation}
\begin{split}
P(s'|s,a)=
\left\{
\begin{array}{l}
{\frac{(M+1) \bar\lambda_{t}}{\sigma(s, a)}, s^{\prime}=\left(M+1, n_{1}, \ldots, n_{N}, A\right)}\\
{\frac{n_{i} i \mu_{t}}{\sigma(s, a)}, \quad s^{\prime}=\left(M+1, n_{1}, \ldots, n_{N}, D_{i}\right)}\\
{\frac{\lambda_{f}}{\sigma(s, a)}, \quad s^{\prime}=\left(M+1, n_{1}, \ldots, n_{N}, F_{+1}\right)}\\
{\frac{\mu_{f}}{\sigma(s, a)}, \quad s^{\prime}=\left(M+1, n_{1}, \ldots, n_{N}, F_{-1}\right)}
\end{array}
\right.
\end{split}
\label{eq27}
\end{equation}

\subsubsection{$s=\left(M, n_{1}, \ldots, n_{N}, F_{-1}\right), a=-1$}

\begin{equation}
\begin{split}
P(s'|s,a)=
\left\{
\begin{array}{l}
{\frac{(M-1) \bar\lambda_{t}}{\sigma(s, a)}, s^{\prime}=\left(M-1, n_{1}, \ldots, n_{N}, A\right)}\\
{\frac{n_{i} i \mu_{t}}{\sigma(s, a)}, \quad s^{\prime}=\left(M-1, n_{1}, \ldots, n_{N}, D_{i}\right)}\\
{\frac{\lambda_{f}}{\sigma(s, a)}, \quad s^{\prime}=\left(M-1, n_{1}, \ldots, n_{N}, F_{+1}\right)}\\
{\frac{\mu_{f}}{\sigma(s, a)}, \quad s^{\prime}=\left(M-1, n_{1}, \ldots, n_{N}, F_{-1}\right)}
\end{array}
\right.
\end{split}
\label{eq28}
\end{equation}

\begin{figure}
\centering
\includegraphics[scale=0.53]{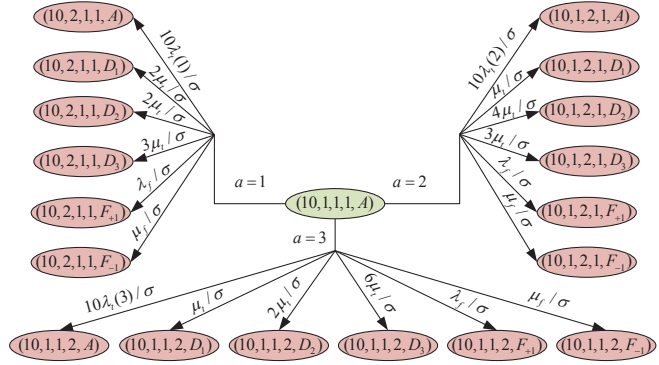}
\caption{State transition diagram.}
\label{fig3}
\end{figure}

A state transition diagram is illustrated in Fig. \ref{fig3} to describe the process of the state transition. In this figure, the current state $s=(10,1,1,1,A)$ and the current event is the arrival of a task. If the action is $1$, $2$ or $3$, the current state will be transmitted to another state with the transition probability as shown in Eq. (\ref{eq25}).

\section{Solution}
\label{sec5}
 In this section, to solve the above problem, we utilize an iteration algorithm to maximize the long-term reward of the SMDP. In each iteration, the maximum value function of each state under different actions is calculated according to the Bellman equation iteratively. The above step repeats until the maximum value function of each state converges. For ease of understanding, we will further describe the proposed algorithm in the following.

\begin{algorithm}
  \caption{Value Iteration Algorithm}
  \label{al1}
  \KwIn{system state set $S$, action set $A_c$, system reward $R(s,a)$, transition probability $P(s'|s,a)$, convergence rate $\epsilon$}
  \KwOut{the optimal scheme $\pi^{*}$}
  Initialization: set $V(s)=0$ for all $s \in S$, and set $k=0$\;
  \For{each system state $s \in S$}
  {
    $\hat{V}_{k+1}(s)=\underset{a \in A}{\max}\left[\hat{R}(s, a)+\hat\gamma \underset{s' \in S}{\sum} \hat{P}\left(s^{\prime} | s, a\right) \hat{V}_{k}\left(s^{\prime}\right)\right]$\;
  }
  \If {$\|\hat{V}_{k+1}-\hat{V}_{k}\| < \epsilon$}
  {
    \For{each system state $s \in S$}
    {
      $\pi^{*}(s)=\underset{a \in A}{\arg \max}\left[\hat{R}(s, a)+\hat\gamma \underset{s' \in S}{\sum} \hat{P}\left(s^{\prime} | s, a\right) \hat{V}_{k+1}\left(s^{\prime}\right)\right]$\;
    }
    }
    \Else {$k++$\; go back to Line 2;}
    {
    }
  Return the optimal scheme $\pi^{*}$\;
\end{algorithm}

 Initially, the number of iteration is set to zero and the value function of each state is initialized to zero. In an iteration,  the maximum value function of each state is calculated according to the Bellman optimal equation \cite{Puterman2005} (as shown in Eq. (\ref{eq29})), based on the initialized value function, rewards and transition probabilities in the previous iteration. For example, in the $k+1th$ iteration, the maximum value function of each state is calculated according to the maximum value function of the $kth$ iteration, rewards and transition probabilities.

 \begin{equation}
V_{k+1}(s)=\max _{a \in A_c}\left[R(s, a)+\gamma \sum_{s^{\prime} \in S} P\left(s^{\prime} | s, a\right) V_{k}\left(s^{\prime}\right)\right],
\label{eq29}
\end{equation}
 Here, $\gamma$ is the discount factor which is used to discount the value function of the next state $s^{\prime}$ and $\gamma=\sigma(s, a) /(\alpha+\sigma(s, a))$.

Then the reward, transition probability and discount factor are normalized to transform the continuous-time SMDP into a discrete-time SMDP \cite{Puterman2005}. The normalized equations are shown as follows,

\begin{equation}
 \hat {R}(s, a)=R(s, a) \frac{\alpha+\sigma(s, a)}{\alpha+y}
 \label{eq30}
\end{equation}
\begin{equation}
\hat {\gamma}=\frac{y}{(y+\alpha)}
\label{eq31}
\end{equation}
\begin{equation}
\hat{P}\left(s^{\prime} | s, a\right)=\left\{\begin{array}{ll}{1-\frac{[1-P(s | s, a)] \sigma(s, a)}{y},} & {s^{\prime}=s} \\ {\frac{P\left(s^{\prime} | s, a\right) \sigma(s, a)}{y},} & {s^{\prime} \neq s}\end{array}\right.
\label{eq32}
\end{equation}
where $y=K \cdot \lambda_{t}+K \cdot N \cdot \mu_{t}+\lambda_{f}+\mu_{f}$. Here $y$ is a normalized factor, which is greater than the largest total event rate.

After that, the Bellman optimal equation can be rewritten as,
\begin{equation}
\hat{V}_{k+1}(s)=\max _{a \in A_c}\left[\hat{R}(s, a)+\hat{\gamma} \sum_{s^{\prime} \in S} \hat{P}\left(s^{\prime} | s, a\right) \hat{V}_{k}\left(s^{\prime}\right)\right]
\label{eq33}
\end{equation}

The normalized maximum value function of each state is calculated according to Eq. (\ref{eq33}) in the $k+1 th$ iteration. After obtaining the maximum value function of each state in the $k+1th$ iteration, the absolute difference of the maximum value function between consecutive iterations is calculated for each state. If the maximum absolute value $\|\hat{V}_{k+1}-\hat{V}_{k}\|$ is smaller than a threshold $\epsilon$, i.e.,

\begin{equation}
\epsilon = \frac{\varepsilon(1-\hat{\gamma})}{2 \hat{\gamma}},
\label{eq34}
\end{equation}
the value iteration algorithm is stopped and the optimal scheme $\pi^{*}$ is the set of the actions corresponding to the maximum value function of each state, i.e., the action of state $s$ in the  optimal scheme $\pi^{*}$ is
\begin{equation}
\pi^{*}(s)=\underset{a \in A_c}{\arg \max}\left[\hat{R}(s, a)+\hat\gamma \underset{s' \in S}{\sum} \hat{P}\left(s^{\prime} | s, a\right) \hat{V}_{k+1}\left(s^{\prime}\right)\right].
\label{eq35}
\end{equation}
Otherwise, the number of the iteration is increased by $1$, then the algorithm continues the next iteration until the optimal scheme is found.

The pseudocode of the iteration algorithm is shown in Algorithm \ref{al1}.

\section{Numerical Results and Analysis}
\label{sec6}

 In this section, we conduct experiments to verify the performance of the optimal scheme through numerical results. The experiment tool is MATLAB 2014b and scenario is described in Section III. In the experiments, we first initialize the tuple of SMDP including the system state set $S$, action set $A_c$, system reward $R(s,a)$ according to Eqs. (\ref{eq1})-(\ref{eq3}), respectively, and transition probability $P(s|s,a)$ according to Eqs. (\ref{eq25})-(\ref{eq28}). Then, we utilize the value iteration algorithm described in Section V to obtain the optimal scheme through maximizing the long-term reward of the SMDP. Finally, we compare the performance of the proposed algorithm with that of the greedy algorithm (GA), which always selects the maximum number of available resources to process the offloaded task \cite{GA}. The GA algorithm tends to reach the local optimization in each step, and thus is suitable to validate the performance of our proposed scheme.
 It is assumed that each task can be offloaded to $3$ RUs at most, i.e., $N=3$. In the numerical results, action $1$, action $2$ and action $3$ indicate that a task is offloaded to one, two and three RUs, respectively; action $0$ indicates the VFC system drops the task. The remaining parameters in the simulation are shown in Table \ref{tab2}.

\begin{table}
\caption{Parameter values of the VFC system.}
\label{tab2}
\footnotesize
\centering
\begin{tabular}{|c|c|c|c|}
\hline
\textbf{Parameter} &\textbf{Value} &\textbf{Parameter} &\textbf{Value}\\
\hline
$N$ &3 &$K$ &5-12\\
\hline
$\lambda_{f}$ &10 &$\mu_{f}$ &10\\
\hline
$\mu_{t}$ &25 / 50 $task/s$ &$\beta $ &5\\
\hline
$T$ &100 $ms$ &$\xi $ &10\\
\hline
$\eta $ &18 &$\alpha $ &0.1\\
\hline
$W_{min}$ &3 &$m$ &1\\
\hline
DIFS &50 $\mu s$ &SIFS &10 $\mu s$\\
\hline
$H$ &229 $\mu s$ &$E[P]$ &1920 bytes\\
\hline
ACK &304 $\mu s$ &ACKtimeout &356 $\mu s$\\
\hline
$slot$ &20 $\mu s$ &$\delta $ &2 $\mu s$\\
\hline
$\varepsilon$ &$10$ &  &\\
\hline
\end{tabular}
\end{table}

\begin{figure}
\centering
\includegraphics[scale=0.6]{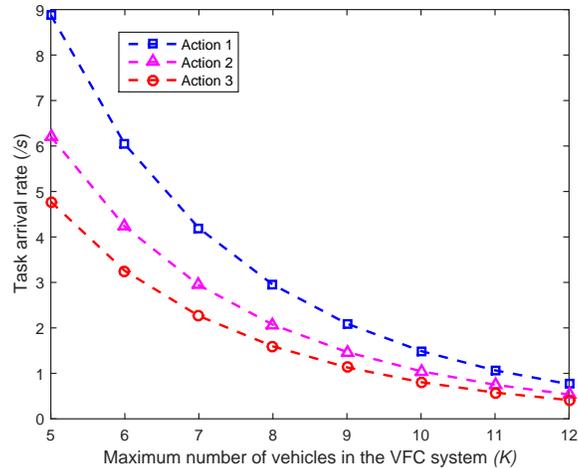}
\caption{Task arrival rate for different maximum number of vehicles in the VFC system.}
\label{fig4}
\end{figure}

 Fig. \ref{fig4} shows the relationships between the task arrival rate and the  different maximum number of vehicles under different actions. We can see that the task arrival rate decreases as the maximum number of vehicles increases. This is because that the collision probability is increased with the increasing maximum number of vehicles, and thus degrades the transmission delay. Thus, the task arrival rate is inverse to the transmission delay. Moreover, it can be seen that the task arrival rate decreases as the number of the allocated RUs increases given the maximum number of vehicle. This is because that number of the subtasks is increased as the number of the allocated RUs increases, thus increasing the number of transmissions. Therefore, with more allocated RUs, the transmission delay is larger and the task arrival rate is smaller.
 
\begin{figure}
\centering
\includegraphics[scale=0.6]{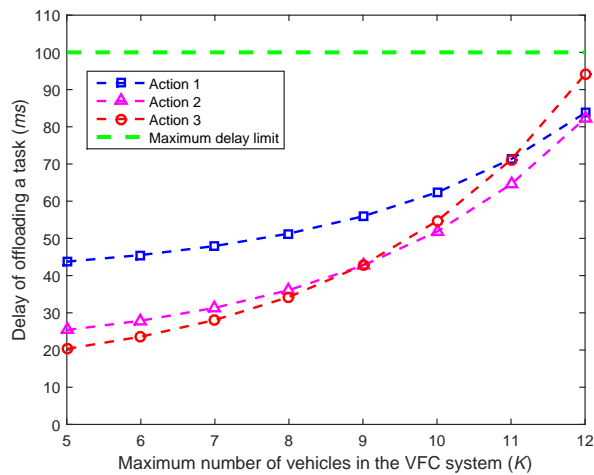}
\caption{Delay of offloading a task for different maximum number of vehicles in the VFC system ($\mu_{t}=25task/s$).}
\label{fig5}
\end{figure}

 Fig. \ref{fig5} shows the relationships between the delay of offloading a task and the maximum number of vehicles in the VFC system under different actions (service rate $\mu_t = 25task/s$). It can be seen that the delay is increased with the increasing maximum number of vehicles. This is because that more vehicles have tasks to offload when the maximum number of vehicles of VFC system increases, thus incurring more collision. In addition, it can be seen that when the maximum number of vehicles is less than $12$, the maximum delay under different actions is less than $100ms$, thus the range of the maximum number of vehicles is set to be no more than $12$ to meet the application requirement. Moreover, it can be seen that when the maximum number of vehicles is small, the delay under action $3$ is smaller than those of other actions. This is because that the number of the request vehicles is small in this case, thus reducing the collision probability and the transmission delay. Moreover, when more RUs are allocated to a task, the computing delay  will be reduced, and thus improves the task offloading delay. However, when the maximum number of vehicles is large, the collision probability is significant, and thus degrade the transmission delay. In this case, although more RUs are allocated, the degraded transmission delay may impact the task offloading delay.

\begin{figure}
\centering
\includegraphics[scale=0.6]{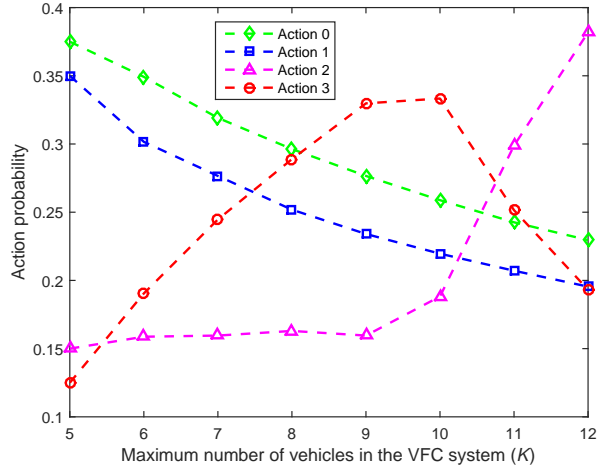}
\caption{Action probability for different maximum number of vehicles in the VFC system ($\mu_{t}=25task/s$).}
\label{fig6}
\end{figure}

 Fig. \ref{fig6} shows the relationships between the action probability of the system and maximum number of vehicles in the VFC system under different actions when the service rate is 25$task/s$. It can be seen that the probability of action $0$ is decreased with the maximum number of vehicles increasing. This is because that when the maximum number of vehicles increases, the number of available RUs in the system also increases, reducing the probability of action $0$. In addition, the system allocates as less RUs as possible when the maximum number of vehicles is small. This is because that the number of available RUs is small in this case, thus the system makes conservative decisions to avoid the case of insufficient available RUs in the system. When the maximum number of vehicles increases, the system is inclined to allocate as more RUs as possible. This is because that with sufficient available RUs, the system makes active decisions to improve long-term rewards. Moreover, action $2$ becomes the best choice when the maximum number of vehicles further increases. This is because that in this case more vehicles transmit task simultaneously, thus deteriorating the collisions and transmission delay. Therefore, action $3$ degrades the transmission delay. On the other hand, action $1$ degrades the computing delay. Thus, action $2$ becomes the best choice in this case to decrease the task offloading delay and get more long-term rewards.

\begin{figure}
\centering
\includegraphics[scale=0.6]{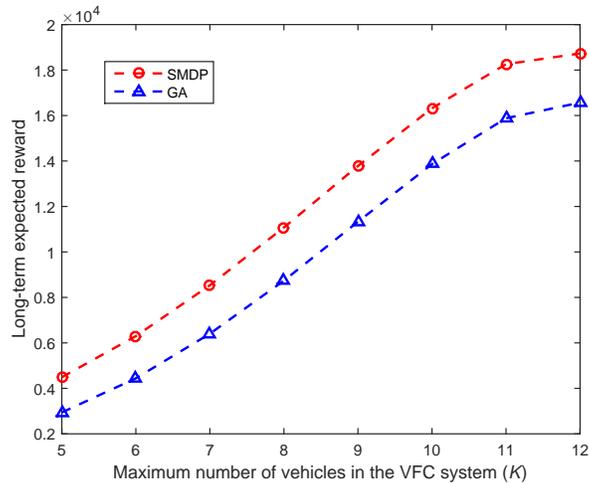}
\caption{Long-term expected reward for different maximum number of vehicles in the VFC system ($\mu_{t}=25task/s$).}
\label{fig7}
\end{figure}

 Fig. \ref{fig7} compares the long-term expected reward of the proposed algorithm and the GA-based scheme when the computation service rate is $25task/s$. It can be seen that the long-term expected reward increases with the increasing maximum number of vehicles. The long-term expected reward the SMDP-based optimal scheme increases by $27.74\%$ as compared with the GA-based scheme. This is attributed to the fact that the GA-based scheme always allocates as many RUs as possible without considering the long-term rewards.

\begin{figure}
\centering
\includegraphics[scale=0.6]{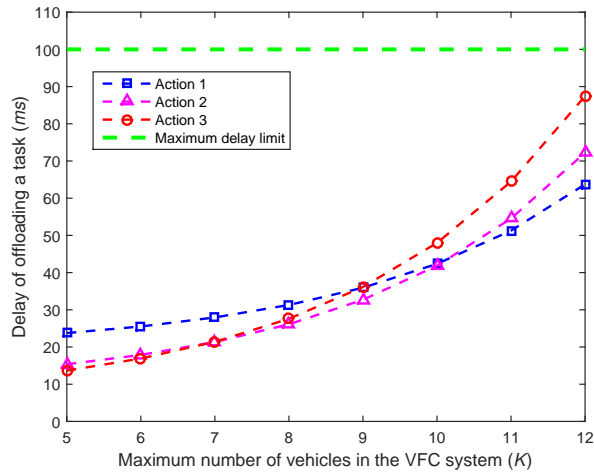}
\caption{Delay of offloading a task for different maximum number of vehicles in the VFC system ($\mu_{t}=50task/s$).}
\label{fig8}
\end{figure}

\begin{figure}
\centering
\includegraphics[scale=0.6]{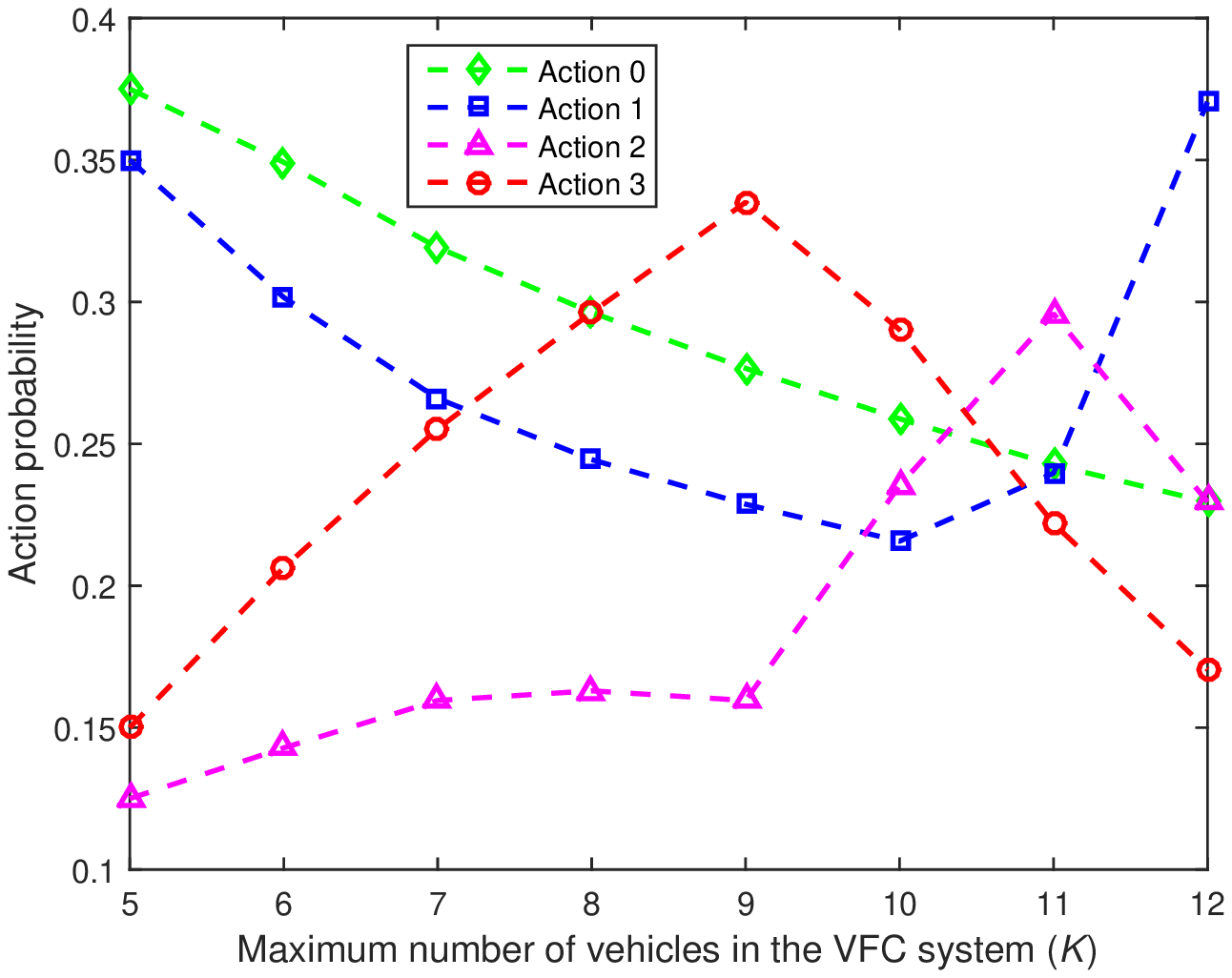}
\caption{Action probability for different maximum number of vehicles in the VFC system ($\mu_{t}=50task/s$).}
\label{fig9}
\end{figure}

\begin{figure}
\centering
\includegraphics[scale=0.6]{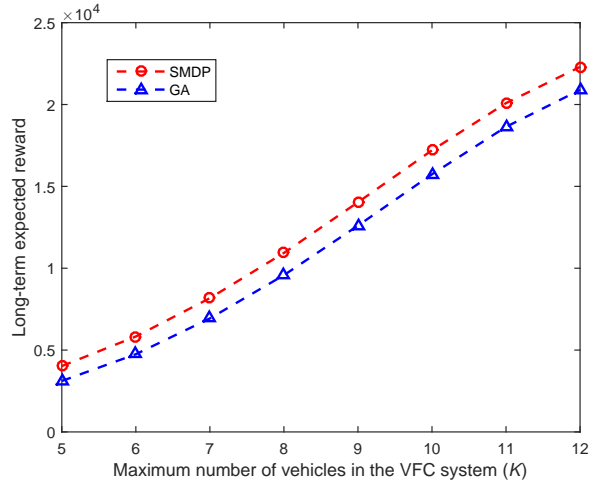}
\caption{Long-term expected reward for different maximum number of vehicles in the VFC system ($\mu_{t}=50task/s$).}
\label{fig10}
\end{figure}

Fig. \ref{fig8} - Fig. \ref{fig10} show the relationships between delay of offloading a task, action probability, long-term expected rewards and the maximum number of vehicles in the VFC system when the service rate $\mu_t=50 task/s$, respectively. In Fig. \ref{fig8}, the trends of the task offloading delay under different actions are the same with the trends in Fig. \ref{fig5}. It can be seen that the maximum number of vehicles in the VFC system is less than $12$ and the maximum delay under different actions is lower than $100ms$. Thus the maximum number of vehicles is also set to be no more than 12 to meet the application requirement when $\mu_t=50task/s$. In Fig. \ref{fig9}, the trends of the action probabilities are almost the same with the trends in Fig. \ref{fig6} when the maximum number of vehicles is no more than $11$. Different from Fig. \ref{fig6}, in Fig. \ref{fig9} when the maximum number of vehicles is increased to $12$, the probability of action $2$ decreases and action $1$ becomes the best choice. This is because that when $\mu_t=50task/s$ the computing delay is much lower as compared with the transmission delay. When maximum number of vehicles increases to $12$, the collisions happens frequently and thus significantly degrades the transmission delay, even if more RUs are allocated. Therefore, the system makes conservative decisions to decrease the transmission delay and improves long-term reward. In Fig. \ref{fig10}, the trends of the action probabilities are the same with the trends in Fig. \ref{fig7} and  the SMDP-based optimal scheme also outperforms the GA-based scheme in terms of the long-term reward. In Fig. \ref{fig10}, the average improvement ratio by using the SMDP-based optimal scheme is $14.91\%$. It can be seen that the improvement ratio in Fig. \ref{fig10} is less than that of Fig. \ref{fig7}. This is because that the allocated RUs process the offloaded tasks faster with $\mu_t=50task/s$ as compared to $\mu_t=25task/s$, thus the average number of the available RUs increases. In this case, the system tries to allocate more RUs as compared with the case when $\mu_t=25task/s$. Therefore, the performance of SMDP-based optimal scheme are much better than the GA-based scheme when $\mu_t=25task/s$ compared with the case when $\mu_t=50task/s$.

\section{Conclusions and Future Work}
\label{sec7}
In this paper, we designed an SMDP model to formulate the task offloading problem in the VFC system where the transmission delay, computing delay, available RUs and the variability feature of vehicles and tasks are jointly taken into account. Then the optimal scheme to maximize the long-term reward was obtained through an iteration method according to Bellman equation. The performance of SMDP-based optimal scheme has been demonstrated by extensive numerical results. Moreover, the long-term expected reward has been verified to be improved by $27.74\%$ and $14.91\%$ when the service rate of a RU is $25$ and $50task/s$, respectively. In the future work, we will  consider the heterogeneity of vehicles and tasks in the task offloading problem of the VFC system.

\ifCLASSOPTIONcaptionsoff
  \newpage
\fi

\bibliographystyle{IEEEtran}
\bibliography{SMDP}

\end{spacing}
\end{document}